\documentclass[a4,aps,twocolumn,showpacs,amsmath,floatfix]{revtex4}
\usepackage{graphicx}%
\usepackage{dcolumn}
\usepackage{color}
\usepackage{amsmath}
\usepackage{here}

\begin{document}
\newcommand{\be}{\begin{equation}}
\newcommand{\ee}{\end{equation}}
\newcommand{\rojo}[1]{\textcolor{red}{#1}}

\title{Two-dimensional nonlinear vector  states in Bose-Einstein condensates}

\author{A. I. Yakimenko$^{1,2}$, Yu. A. Zaliznyak$^1$, and V. M. Lashkin$^{1}$}

\affiliation{$^1$Institute for Nuclear Research, Kiev 03680, Ukraine \\
$^2$Department of Physics, Taras Shevchenko National University,
Kiev 03022, Ukraine}

\begin{abstract}
Two-dimensional (2D) vector matter waves in the form of
soliton-vortex and vortex-vortex pairs are investigated for the
case of attractive intracomponent interaction in two-component
Bose-Einstein condensates. Both attractive and repulsive
intercomponent interactions are considered. By means of a linear
stability analysis we show that soliton-vortex pairs can be stable
in some regions of parameters while vortex-vortex pairs turn out
to be always unstable. The results are confirmed by direct
numerical simulations of the 2D coupled Gross-Pitaevskii
equations.
\end{abstract}


\maketitle

\section{Introduction}

Multicomponent Bose-Einstein condensates (BECs) have been subject
of growing interest in recent years as they open intriguing
possibilities for a number of important physical applications,
including coherent storage and processing of optical fields
\cite{op1,op2}, quantum simulation \cite{qubit}, quantum
interferometry \textit{etc}. Experimentally, multicomponent BECs
can be realized by simultaneous trapping of different species of
atoms \cite{DifAtom1,DifAtom2} or  atoms of the same isotope in
different hyperfine states. Magnetic trapping freezes spin
dynamics \cite{MagTrap1,MagTrap2}, while in optical dipole traps
all hyperfine states are liberated (spinor BECs) \cite{OpTrap1}.
Theoretical models of multicomponent BECs in the mean-field
approximation are formulated in the framework of coupled
Gross-Pitaevskii (GP) equations \cite{Dalfovo} and the order
parameter of multicomponent BECs  is described by a multicomponent
vector.

Like in the scalar condensate case, various types of nonlinear
matter waves have been predicted in multicomponent BECs. They
include, in addition to ground-state solutions
\cite{ground1,ground2,ground3}, structures which are peculiar to
 multicomponent BECs only, such as bound states of dark-bright
\cite{darkbright} and dark-dark \cite{darkdark}, dark-gray,
bright-gray, bright-antidark and dark-antidark \cite{darkgrey}
complexes of solitary waves, domain wall solitons
\cite{wall1,wall2,wall3}, soliton molecules \cite{Molecul},
symbiotic solitons \cite{symbiotic}. Two-dimensional (2D) and
three-dimensional (3D) vector solitons and vortices have been
considered in Refs. \cite{Skryabin,Garcia1,Garcia2} for the case
of repulsive condensates. Attractive intracomponent interaction
have, on the other hand, received less attention and only
one-dimensional vector structures have been studied so far
\cite{attract1D,Dutton}. Two-dimensional and 3D cases, however,
demands special attention since the phenomenon of collapse is
possible in attractive BECs.

Interactions between the atoms in the same and different states
can be controlled (including changing the sign of the
interactions) via a Feshbach resonance. Theoretical and
experimental studies have shown that inter-component interaction
plays a crucial role in dynamics of nonlinear structures in
multicomponent BECs. Recently, two-component BECs with tunable
inter-component interaction were realized experimentally
\cite{exp1,exp2}. Note that in nonlinear optics, where similar
model equations (without the trapping potential) are used to
describe the soliton-induced waveguides \cite{Kivshar}, the
nonlinear coefficients are always of the same sign.

The aim of this paper is to study 2D nonlinear localized vector
structures in the form of soliton-vortex and vortex-vortex pairs
in a binary mixture of disc-shaped BECs with attractive
intracomponent and attractive or repulsive intercomponent
interactions. Then, by means of a linear stability analysis, we
investigate the stability of these structures and show that pairs
of soliton and single-charged vortex  can be stable both for
attractive and repulsive interactions between different
components. Vortex-vortex pairs turn out to be always unstable.
The results are confirmed by direct numerical simulations of the
2D coupled Gross-Pitaevskii equations.

The paper is organized as follows. In Sec. ~\ref{sec2} we
formulate a model and present basic equations. The cases of
attractive and repulsive intercomponent interactions are
considered in Secs. ~\ref{sec3} and ~\ref{sec4} respectively. The
conclusions are made in Sec. ~\ref{sec5}.

\section{Basic equations}
\label{sec2}
 We consider a binary mixture of BECs, consisting of two
different spin states of the same isotope. We assume that the
nonlinear interactions are weak relative to the confinement in the
longitudinal (along $z$-axis) direction. In this case, the BEC is
a "disk-shaped" one, and the GP equations take an effectively 2D
form
\begin{equation}
\label{GP1}
 i\hbar\frac{\partial \Psi_1}{\partial t}=\left[-\frac{\hbar^2}{2M}\nabla^2
 +V_\mathrm{ext}(\mathbf{r})+g_{11}|\Psi_1|^2+g_{12}|\Psi_2|^2\right]\Psi_1,
\end{equation}
\begin{equation}
\label{GP2}
 i\hbar\frac{\partial \Psi_2}{\partial t}=\left[-\frac{\hbar^2}{2M}\nabla^2
  +V_\mathrm{ext}(\mathbf{r})+g_{21}|\Psi_1|^2+g_{22}|\Psi_2|^2\right]\Psi_2,
\end{equation}
where $M$ is the mass of the atoms,
$V_\mathrm{ext}(\mathbf{r})=M\omega_{\perp}^{2}(x^{2}+y^{2})/2$ is
the harmonic external trapping potential with frequency
$\omega_\perp$ and $\nabla^2=\partial^{2}/\partial
x^{2}+\partial^{2}/\partial y^{2}$ is the 2D Laplacian. Atom-atom
interactions are characterized by the coupling coefficients
$g_{ij}=4\pi\hbar^2 a_{ij}/M$, where $a_{ij}=a_{ji}$ are the
$s$-wave scattering lengths for binary collisions between atoms in
internal states $|i\rangle$ and $|j\rangle$. Note that
$g_{11}=g_{22}$ and $g_{12}=g_{21}$. Introducing dimensionless
variables $(x,y)\to(x/l,y/l)$, $t\to \omega_{\perp} t$,
$\Psi_j\to\Psi_j\sqrt{\hbar \omega_\perp/(2|g_{11}|)}$,
$B_{12}=-g_{12}/|g_{11}|$, $B_{11}=-g_{11}/|g_{11}|$, where
$l=\sqrt{\hbar/(M\omega_{\perp})}$, one can rewrite equations
(\ref{GP1}) and (\ref{GP2}) as

\begin{equation}
\label{main1}
 i\frac{\partial \Psi_1}{\partial t}=\left[-\nabla^2
 +x^{2}+y^{2}-|\Psi_1|^2-B_{12}|\Psi_2|^2\right]\Psi_1,
\end{equation}
\begin{equation}
\label{main2}
 i\frac{\partial \Psi_2}{\partial t}=\left[-\nabla^2
  +x^{2}+y^{2}-B_{12}|\Psi_1|^2-|\Psi_2|^2\right]\Psi_2.
\end{equation}
In what follows we consider attractive interaction between atoms
of the same species and set $B_{11}=B_{22}=1$. We neglect the spin
dynamics (assuming magnetic trapping) so that the interaction
conserves the total number $N_{j}$ ($j=1,2$) of particles of each
component
\begin{equation}
\label{N} N_j=\int{|\Psi_j|^2d^2\textbf{r}},
\end{equation}
and energy
\begin{equation}
\label{E} E=E_1+E_2-\frac12
B_{12}\int{|\Psi_1|^2|\Psi_2|^2d^2\textbf{r}},
\end{equation}
where
\begin{equation}
E_j=\int\left\{|\nabla\Psi_j|^2-
\frac12|\Psi_j|^4+(x^{2}+y^{2})|\Psi_j|^2\right\} d^2\textbf{r}.
\end{equation}

\section{Attractive intercomponent interaction}
\label{sec3}
\subsection{Stationary soliton-vortex pairs}
We look for stationary solutions of Eqs. (\ref{main1}) and
(\ref{main2}) in the form
\begin{equation}
\label{Psi} \Psi_j(\mathbf{r},t)=\psi_j(r)e^{-i\mu_j t+i
m_j\varphi}
\end{equation}
where $m_j$ is the topological charge (vorticity) of the $j$-th
component, $\mu_j$ is the chemical potential, $r=\sqrt{x^2+y^2}$
and $\varphi$ is the polar angle. Substituting Eq. (\ref{Psi})
into Eqs. (\ref{main1}) and (\ref{main2}), we have
\begin{equation}
\label{stat1}
\mu_1\psi_1+\Delta_r^{(m_1)}\psi_1-r^2\psi_1+(|\psi_1|^2+B_{12}|\psi_2|^2)\psi_1=0,
\end{equation}
\begin{equation}
\label{stat2}
\mu_2\psi_2+\Delta_r^{(m_2)}\psi_2-r^2\psi_2+(B_{12}|\psi_1|^2+|\psi_2|^2)\psi_2=0,
\end{equation}
where $\Delta_r^{(m)}=d^2/d r^2+(1/r)(d/d r)-m^2/r^2$. As was
pointed out the inter-component interaction may be varied over
wide range, however, the strength of the intercomponent
interaction is weaker than the intracomponent counterpart in most
experiments with two-component BECs. Thus $B_{12}$ can be
considered as the free parameter from the range $-1\le B_{12}\le
1$. We find that the qualitative behavior of vector state
characteristics does not change when varying $B_{12}$. To make it
definite we further fixed the strength of intercomponent
interaction at $B_{12}=\pm0.5$ for attractive and repulsive cases
respectively. In this section we consider the case of attractive
interatomic interaction $B_{12}>0$.
\begin{figure}
\includegraphics[width=3.4in]{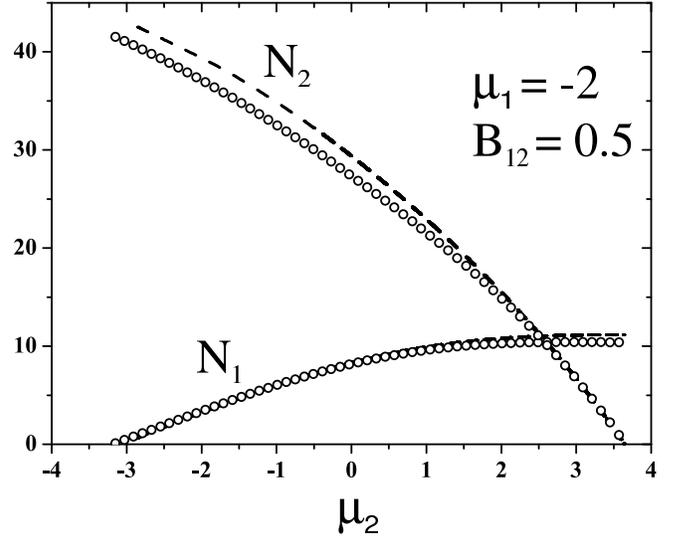}
\caption{Normalized numbers of atoms $N_{1}$ and $N_{2}$ of each
component versus chemical potential $\mu_2$ at fixed $\mu_1$ for
vector soliton-vortex pair ($m_1=0$, $m_2=1$) (attractive
intercomponent interaction).} \label{EDDS_N1N2}
\end{figure}
\begin{figure}
\includegraphics[width=3.4in]{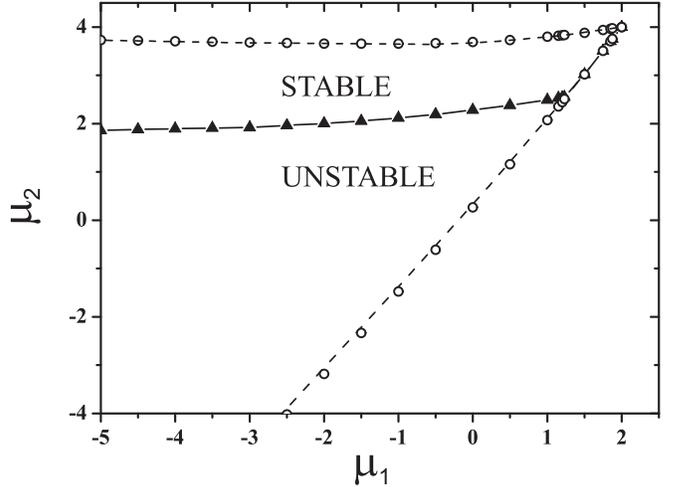}
\caption{Existence domain for vector state $m_1=0$ and $m_2=1$ on
the $(\mu_1,\mu_2)$ plane for attractive intercomponent
interaction $B_{12}=0.5$. Open circles correspond to numerically
found existence boundaries. Dashed curves outline the variational
predictions. At upper and lower boundaries of the existence domain
the vector states degenerate into the pure scalar states with
$N_1=0$ and $N_2=0$ respectively. The solid line with triangles
indicates the stability boundary.} \label{Figure1}
\end{figure}
\begin{figure*}
\includegraphics[width=6.8in]{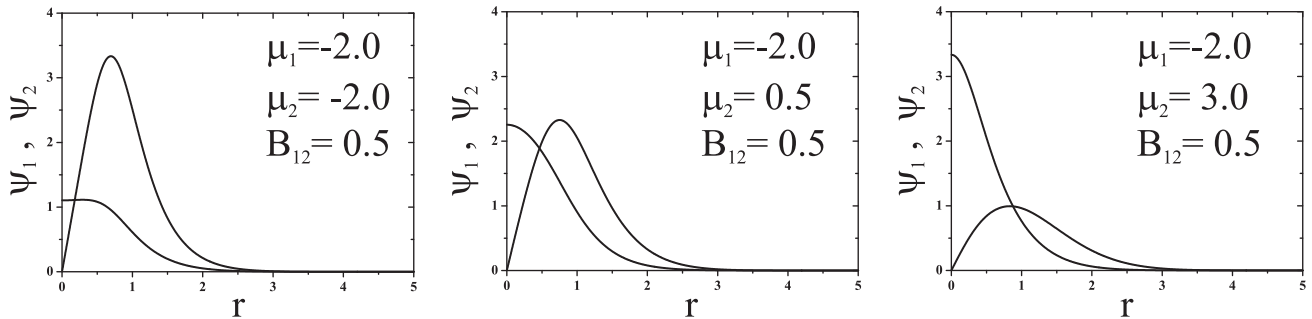}
\caption{The numerically found solutions of Eqs. (\ref{stat1}) and
(\ref{stat2}) for attractive intercomponent interactions
$B_{12}>0$ at fixed $\mu_1=-2$ for different values of $\mu_2$.}
\label{Figure2}
\end{figure*}
\begin{figure}
\includegraphics[width=3.4in]{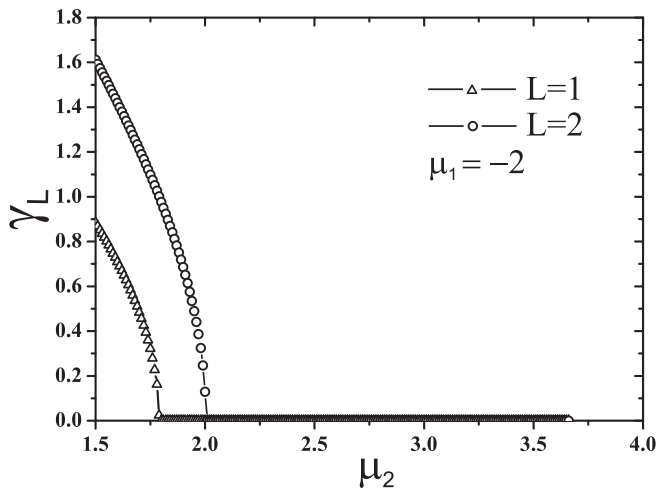}
\caption{Typical dependence of maximum growth rates for $L=1$ and
$L=2$ azimuthal modes as functions of $\mu_2$ at fixed $\mu_1$,
(here $\mu_1=-2$), $B_{12}=0.5$. Note that the widest instability
domain has $L=2$ mode.} \label{Figure3}
\end{figure}
\begin{figure}
\includegraphics[width=3.4in]{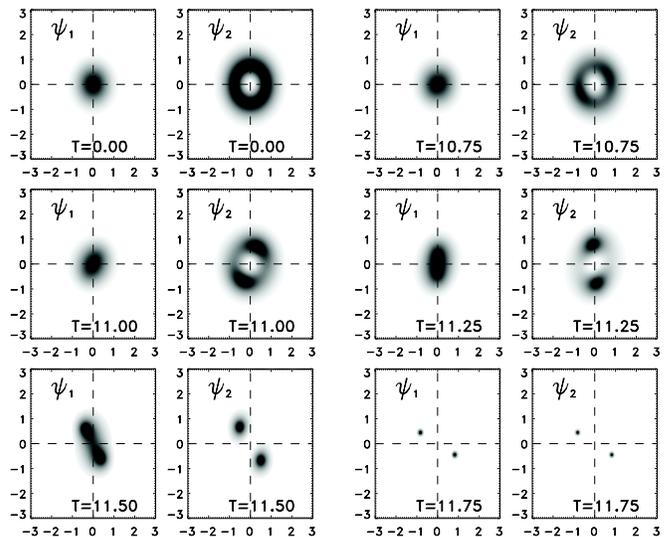}
\caption{Snapshots of unstable evolution for vector pair $m_1=0$,
$m_2=1$, $\mu_1=-2$, $\mu_2=0.5$, $B_{12} = 0.5$. The absolute
values of $|\psi_1|$ and $|\psi_2|$ are shown in grayscale: the
darker regions correspond to higher amplitudes.}
\label{Evolution1}
\end{figure}
First, we present a variational analysis. Stationary solutions of
Eqs. (\ref{stat1}) and (\ref{stat2}) in the form Eq. (\ref{Psi})
realize the extremum of the energy functional $E$ under the fixed
number of particles $N_{1}$ and $N_{2}$. We take trial functions
$\psi_{j}$ in the form
\begin{equation}\label{trial}
\psi_j(r)=h_j\left(\frac{r}{a_j}\right)^{|m_j|}\exp{\left(-
\frac{r^2}{2a_j^2}+im_j\varphi\right)},
\end{equation}
which correspond to the localized state $(m_{1},m_{2})$ with
vorticities $m_{1}$ and $m_{2}$ for $|1\rangle$ and $|2\rangle$
components respectively, $a_j$ and $h_j$ are unknown parameters to
be determined by the variational procedure. The parameters $h_1$
and $h_2$ can be excluded using the normalization conditions
(\ref{N}), which yield the relation $N_j= m_j!\pi h_j^2a_j^2$.
Thus, the only variational parameters are $a_1$ and $a_2$.
Substituting Eq. (\ref{trial}) into Eq. (\ref{E}), we get for the
functional $E$
\begin{equation}
\label{E}\nonumber E=E_1+E_2+E_{12},
\end{equation}
where
$$E_j=\frac{N_j}{ a_j^2}\left(m_j+1-\frac{N_j^2(2m_j)!}{\pi4^{1+m_j}(m_j!)^2}\right)+N_ja_j^2(m_j+1),$$
and
$$ E_{12}=-\frac12B_{12}N_1N_2\frac{a_1^{2|m_1|}a_2^{2|m_2|}
}{(a_1^2+a_2^2)^{1+|m_1|+|m_2|}}\frac{(m_1+m_2)!}{\pi m_1!m_2!}.
$$
By solving the variational equations $\partial E/\partial a_j=0$
at fixed $N_j$ one finds the parameters of approximate solutions
with different $m_{1}$ and $m_{2}$. We will focus on one
particular configuration with $m_{1}=0$ and $m_{2}=1$ which
corresponds to the pair soliton-vortex. The results of the
variational analysis for this case and $B_{12}>0$ are given in
Fig. \ref{EDDS_N1N2} and Fig. \ref{Figure1} by dashed lines. These
results were the starting point for numerical analysis.

The equations (\ref{stat1}) and (\ref{stat2}) were discretized on
the equidistant radial grid and the resulting system was solved by
the stabilized iterative procedure similar to that described in
Ref. \cite{Petviashvili86}. The appropriate initial guesses were
based on the variational approximation. The numerical results are
shown in Fig. \ref{EDDS_N1N2} and Fig. \ref{Figure1} by open
circles. It is seen that the variational results exhibit a good
agreement with numerical calculations.

The stationary vector states form two-parameter family with
parameters $\mu_1$ and $\mu_2$.
 In the Fig. \ref{EDDS_N1N2} the
number of atoms for each component of the stationary vector state
$(0,1)$ is represented as a function of the chemical potential
$\mu_2$ at fixed value of $\mu_1$. The existence domain is bounded
and its boundaries are determined by the condition
$\left.\mu_2\right|_{N_1=0}\le\mu_2\le\left.\mu_2\right|_{N_2=0}$.
For each value of $\mu_1$, where the solution exists, a dependence
similar to one presented in Fig. \ref{EDDS_N1N2} can be found.
This allows one to reconstruct an existence domain of the vector
pair $(0,1)$ on ($\mu_1$, $\mu_2$) plane, which is shown in the
Fig. \ref{Figure1}. As is known, (see e.g.
\cite{MihalacheMalomedPRA06}) for the two-dimensional scalar
solitary structures in BEC with attraction, the chemical potential
is bounded from above $\mu<\mu_\mathrm{{max}}=2(m+1)$, where $m$
is the topological charge, and $N\to 0$ when
$\mu\to\mu_\mathrm{{max}}$. One can see from Fig. \ref{Figure1}
that the value of $\mu_\mathrm{{max}}$ is reduced in the presence
of the second component if the intercomponent interaction is
attractive ($B_{12}>0$). Both components vanish at the point
$(\mu_1, \mu_2)=(2m_1+2,2m_2+2)$.

Examples of soliton-vortex $(0,1)$ radial profiles are given in
Fig. \ref{Figure2}. It is interesting to note that when the
amplitude of the vortex component is sufficiently high, a soliton
profile develops a noticeable plateau. Such a deviation from
gaussian-like shape leads to comparable divergence of numerical
and variational dependencies in Fig. \ref{EDDS_N1N2}. Other vector
states as $(0,2)$, $(-1,1)$, $(-2,2)$  \textit{etc.} were also
found, but they all turn out to be always unstable (see below).

\begin{figure}[h]
\includegraphics[width=3.4in]{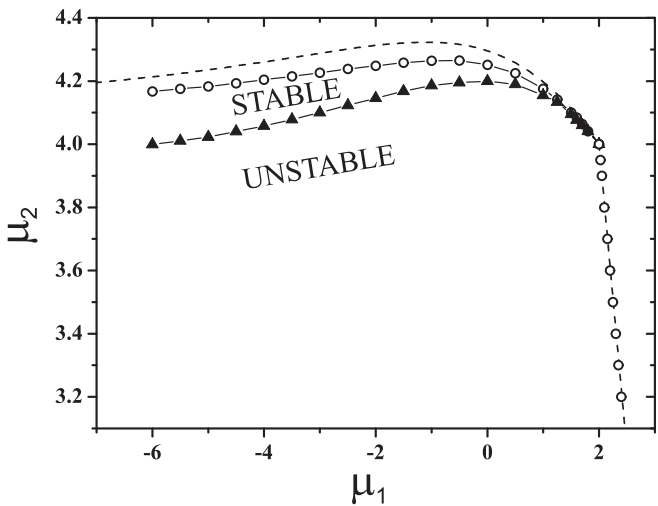}
\caption{The same as in Fig. \ref{Figure1} for repulsive
intercomponent interaction ($B_{12}=-0.5$).} \label{Figure4}
\end{figure}

\begin{figure}[h]
\includegraphics[width=3.4in]{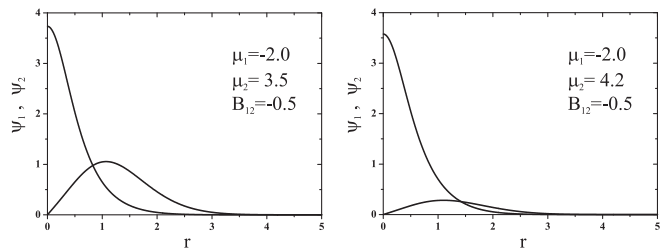}
\caption{Numerically found solutions of Eqs. (\ref{stat1}) and
(\ref{stat2}) for repulsive intercomponent interactions at fixed
$\mu_1=-2$ from stable (right panel) and unstable (left panel)
regions.} \label{Figure5}
\end{figure}

\begin{figure}[h]
\includegraphics[width=3.4in]{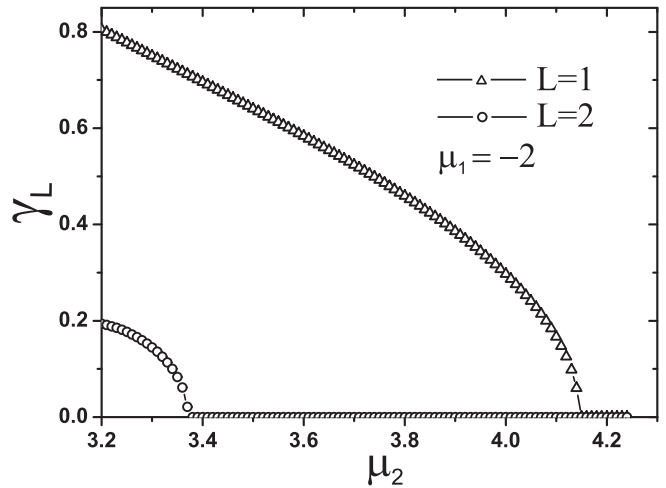}
\caption{Same as in Fig. \ref{Figure3} for $B_{12}=-0.5$. Note
that in contrast to the case of attractive intercomponent
interaction the stability threshold is determined by $L=1$ mode.}
\label{Figure6}
\end{figure}

\begin{figure}
\includegraphics[width=3.4in]{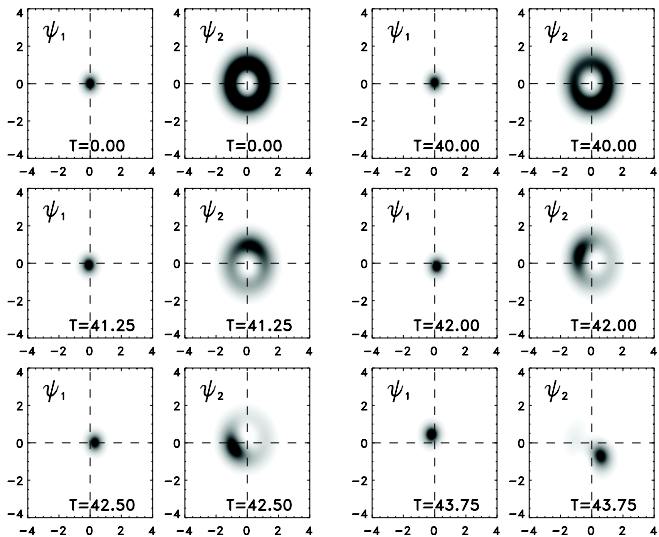}
\caption{Development of snake-type instability ($L=1$) for vector
pair $m_1=0$, $m_2=1$, $\mu_1=-2$, $\mu_2=3.5$. Intercomponent
interaction is repulsive, $B_{12} = -0.5$. The absolute values of
$|\psi_1|$ and $|\psi_2|$ are shown for different times.}
\label{Evolution2}
\end{figure}

\subsection{Stability of stationary solutions}
The stability of the vector pairs can be investigated by the
analysis of small perturbations of the stationary states. We take
the wave functions in the form
\begin{equation}
 \Psi_j(\textbf{r} ,t)=
\left\{\psi_j(r)+\varepsilon_j(\textbf{r} ,t)\right\}e^{-i\mu_j
 t+im_j\varphi},
\end{equation}
where the stationary solutions $\psi_j(r)$ are perturbed by  small
perturbations $\varepsilon_j(\textbf{r} ,t)$, and linearize Eqs.
(\ref{main1}) and (\ref{main2}) with respect to $\varepsilon_j$.
The basic idea of such a linear stability analysis is to represent
a perturbation as the superposition of the modes with different
azimuthal symmetry. Since the perturbations are assumed to be
small, stability of each linear mode can be studied independently.
Presenting the perturbations in the form
 $$\varepsilon_j(\textbf{r} ,t)=u_j(r)e^{i\omega t+i L\varphi}
 +v_j^*(r)e^{-i\omega^* t-i L\varphi},$$
we get the
 following
 linear eigenvalue problem
\begin{equation}\label{eq:lineareigen}
 \left(%
\begin{array}{cccc}
  \hat L_{12}^{(+)} & \alpha_1 & \beta_{12} & \beta_{12} \\
-\alpha_1   & -\hat L_{12}^{(-)} & -\beta_{12} & -\beta_{12} \\
\beta_{12}   & \beta_{12} & \hat L_{21}^{(+)}  & \alpha_2\\
-\beta_{12}   & -\beta_{12} &  \alpha_2 & -\hat L_{21}^{(-)}  \\
\end{array}%
\right) U= \omega U,
\end{equation}
where $U=(u_1,v_1,u_2,v_2)$ is the vector eigenmode and $\omega$
is an (generally, complex) eigenvalue, $\alpha_j=\psi_j^2$,
$\beta_{12}=B_{12}\psi_1\psi_2$, $\hat
L_{ij}^{(\pm)}=\mu_i+\Delta_r^{(m_i\pm
L)}-r^2+2\psi_i^2+B_{ij}\psi_j^2$. An integer $L$ determines the
number of the azimuthal mode. Nonzero imaginary parts in $\omega$
imply the instability of the state $|\psi_{1},\psi_{2}\rangle$
with $\gamma_{L}=\max |\mathrm{Im}\,\omega|$ being the instability
growth rate.

Employing a finite difference approximation, we numerically solved
the eigenvalue problem (\ref{eq:lineareigen}).
Typical dependencies of the growth rate $\gamma_{L}$ of the
azimuthal perturbation modes $L=1$ and $L=2$ on $\mu_2$ at fixed
$\mu_1$ are shown in Fig. \ref{Figure3} for the state $(0,1)$.
Note that above some critical value of $\mu_2$ all growth rates
vanish granting the stability of the vector pair $(0,1)$ against
the azimuthal perturbations. For the case of attractive
intercomponent interaction $B_{12}>0$ the stability boundary is
determined by $L=2$ mode. Similar dependencies of the growth rate
$\gamma_{L}$ on $\mu_2$ have been obtained for other values of
$\mu_1$. We have performed the numerical calculation of
$\gamma_{L}$ for values of the azimuthal index up to $L=5$. In all
studied cases azimuthal stability is always defined by vanishing
of the growth rate of $L=2$ mode. The corresponding stability
threshold is given in Fig. \ref{Figure1} by filled triangles. Note
that in the degenerate scalar case $\psi_{1}=0$ for single-charge
vortex $m_2=1$ the stability threshold $\mu_2=2.552$ coincides
with the value obtained in Ref. \cite{MihalacheMalomedPRA06}. For
the vector states $(0,2)$, $(-1,1)$, $(-2,2)$, the growth rates
are found to be nonzero in the entire existence domain
 and these pairs appear to be always unstable.

To verify the results of the linear analysis, we solved
numerically the dynamical equations (\ref{main1}) and
(\ref{main2}) initialized with our computed vector solutions.
Numerical integration was performed on the rectangular Cartesian
grid with a resolution $512^2$ by means of standard split-step
fourier technique (for details see e.g. \cite{KivsharAgrawal}). In
full agreement with the linear stability analysis, the states
$(0,1)$ perturbed by the azimuthal perturbations with different
$L$ survive over huge times provided that the corresponding
$\mu_{1}$ and $\mu_{2}$ belong to the stability region. On the
other hand, Fig. \ref{Evolution1} shows the temporal development
of azimuthal $L=2$ instability for the vector state $(0,1)$ with
$\mu_{1}=-2$ and $\mu_{2}=0.5$ (i. e. in the instability region).
One can see the two humps which appear on the initially smooth
ring-like intensity distribution. Further, the vortex profile is
deformed, vortex and fundamental soliton both split into two
filaments which then collapse. Note that unstable $(0,1)$ pair in
BECs with repulsive intracomponent interaction \cite{Garcia1} does
not collapse and undergo a complex dynamics with trapping one
component by another.

\section{Repulsive intercomponent interaction}
\label{sec4}

In this section we present results for the case of repulsive
interactions $B_{12}<0$ between different components.
 The existence domain, stable and unstable branches on the $(\mu_1,
\mu_2)$ plane for the state $(0,1)$ are shown in Fig.
\ref{Figure4}. It is seen that the repulsive intercomponent
interaction leads to an increase of the maximum chemical potential
$\mu_\mathrm{{max}}$ for each component compared to the case
$\mu=4$ of pure scalar solution. The stability properties of the
vector states were investigated by the linear stability analysis
described in the preceding section. The states $(0,2)$, $(-1,1)$,
$(-2,2)$ are always unstable as in the case of $B_{12}>0$.

Figure \ref{Figure5} shows examples of the radial profiles of
unstable and stable $(0,1)$ states. In Fig. \ref{Figure6} we plot
the growth rates $\gamma_{L}$  as functions of $\mu_{2}$ under
fixed $\mu_{1}$ for the azimuthal perturbations with $L=1$ and
$L=2$. The growth rates vanish if $\mu_{2}$ exceeds a some
critical value. In contrast to the attractive
 intercomponent interaction case,
it is seen that the stability boundary is controlled by the
elimination of the snake-type instability (i. e. azimuthal
perturbation with $L=1$). Indeed, the repulsion between components
naturally leads to spatial separation of condensate species. This
relative motion destroys the vector state as seen from Fig.
\ref{Evolution2}.

\section{Conclusions}
\label{sec5}

In conclusion, we have analyzed the stability of 2D vector matter
waves in the form of soliton-vortex and vortex-vortex pairs in
two-component Bose-Einstein condensates with attractive
interactions between atoms of the same species. Both attractive
and repulsive intercomponent interactions are considered. We have
performed a linear stability analysis and showed that, in both
cases, only soliton-vortex pairs $(0,1)$ can be stable in some
regions of parameters. Namely, under the fixed number of atoms in
the soliton component, the number of atoms of the vortex component
should be less than a some critical value. No stabilization
regions have been found for vortex-vortex pairs and they turn out
to be always unstable. The results of the linear analysis have
been confirmed by direct numerical simulations of the 2D coupled
Gross-Pitaevskii equations.

\end{document}